\documentclass[eqsecnum,aps,epsfig]{revtex4}
\usepackage{amsmath,amssymb,amsthm,bm}
\usepackage{psfrag}
\usepackage[dvips]{graphicx}
\usepackage[utf8]{inputenc} 
\usepackage{hyperref}

\begin{document}
\title{Domain structures and quark potentials in SU($3$) gauge theory}
\author{Seyed Mohsen Hosseini Nejad}
\email{smhosseininejad@ut.ac.ir}
\affiliation{
Faculty of Physics, Semnan University, P.O. Box 35131-19111, Semnan, Iran}

\begin{abstract}

We analyze the static potentials for various representations in SU($3$) Yang-Mills theory
within the framework of the domain model of center vortices. The influence of vortex interactions is investigated on the static potentials. We show that, by ad-hoc choosing the probability weights of the
different vortex configurations contributing to the static potential, a phenomenologically satisfactory result for
the different representations can be achieved. In particular including vacuum domains, a way to effectively parametrize
vortex interactions, is crucial in obtaining an (almost)
everywhere convex potential when interpolating between the short
distances and the asymptotic regimes.   
\end{abstract}

\maketitle

\section{INTRODUCTION}\label{Sect0}
Understanding quark confinement and the dynamical mechanism behind it is a big challenge in QCD. The interaction between static quark sources at small separations is dominated by one-gluon exchange and the potential
is Coulomb-like. At intermediate distances, quark confinement arises referring to the color electric flux-tube formation and linear potentials. In this range of distances, the string
tensions for different representations are qualitatively in agreement with Casimir scaling \cite{Deldar:1999vi,Bali:2000un,Piccioni:2005un}. At asymptotic distances, the string tensions depend only on the $N$-ality of the representations \cite{Kratochvila:2003zj}. In addition, the quark potential must be everywhere convex and without concavity \cite{Bachas:1985xs}.
Numerical simulations \cite{DelDebbio:1996lih,Langfeld:1997jx,DelDebbio:1997ke,Langfeld:1998cz,Engelhardt:1999fd,Kovacs:1998xm} and infrared models \cite{Faber:1997rp,Greensite:2006sm,Engelhardt:1999wr,Engelhardt:2003wm,Deldar:2010hw,Deldar:2011fh,Deldar:2009aw,Nejad:2014hka} have indicated that center vortices \cite{tHooft:1977nqb,Vinciarelli:1978kp,Yoneya:1978dt,Cornwall:1979hz,Mack:1978rq,Nielsen:1979xu} which are quantized magnetic flux tubes could account for the quark confinement via the area law of the Wilson loop. Furthermore, numerical simulations
have shown that the center vortices could also account for spontaneous chiral symmetry breaking \cite{deForcrand:1999our,Engelhardt:2002qs,Hollwieser:2008tq,Bowman:2010zr,Nejad:2016fcl,Hollwieser:2013xja,Nejad:2018pfl}. 

The thick center vortex model \cite{Faber:1997rp,Greensite:2006sm} is a phenomenological model trying to understand
the color confinement in terms of the interaction of the Wilson loops with the center vortices. However, the potentials induced by center vortices for some representations show unphysical concavity when interpolating between the short distances and the asymptotic regimes. For removing the concavity, in Ref. \cite{Deldar:2010hw}, the vortex profile is allowed to fluctuate. 

In this paper, this artifact is studied through analyzing vortex interactions. We represent various forms of the center vortex picture of confinement for
static quark potentials in different representations of SU($3$). In the thick center vortex model, we investigate the Yang–Mills vacuum of the SU($3$) gauge theory including two types of center vortices. In some literatures, two types of vortices may be regarded as the
same type of vortex but with magnetic flux pointing in opposite directions. Without this constraint, we study the behavior of these center vortices on static potentials in this analytical model. Although interactions of both types of center vortices with sufficiently large Wilson loops are the same, their interactions with medium size Wilson loops are different. Besides, Casimir scaling and $N$-ality regimes for some representations do not connect smoothly and some kind of unexpected concavity occurs in the model which explicitly disagrees with
lattice results. In Refs. \cite{Greensite:2006sm,Liptak:2008gx,Nejad:2014tja}, a domain structure is assumed in the vacuum for G($2$) and SU($N$) gauge theories. The total magnetic flux
through each domain corresponds to a center element of $\mathbb{Z}(N)$ subgroup. In the framework of the domain model of center vortices, we analyze the domain structures with a fixed vortex profile for removing concavity and improving Casimir scaling especially for higher representations of the SU($3$) gauge group. Interactions between two types of vortices are discussed using a dual analogy to the type II superconductivity where it seems that two vortices repel each other while the vortex-antivortex interaction is attractive. Moreover, we argue that the same interactions may be confirmed by the model. We show that interactions between two types of center vortices may deform them to the configurations with the lowest magnitude of center fluxes where there are appeared vortices of type one as well as vacuum domains on the vacuum. We show that ad-hoc choosing the probability weights of the
domain structures is crucial in obtaining an (almost)
everywhere convex potential. 

In Sec. \ref{Sect1}, we analyze the static potentials in various representations and their ratios induced by center vortices in SU($3$) gauge theory within the framework of the thick center vortex model. We investigate the contributions of center vortices and the vortex interactions in the potentials in Sec. \ref{Sect2}. Then, in Sec. \ref{Sect3} the confinement mechanism in the background of center vortices would be reformulated for removing concavity and improving the Casimir scaling. We summarize the main points of our study in Sec. \ref{Sect4}. 

\section{Static potentials induced by two types of SU($3$) center vortices}\label{Sect1}
Any non-Abelian SU($N$) gauge theory of confinement should explain some features of the confining force which can be verified in lattice simulations. If one neglects dynamical quarks in the vacuum in the
first approximation, the static quark potential of nonperturbative regime has distinct behavior in two ranges of interquark distances. At intermediate distances, from the onset of the confinement to the onset of color screening, the quark potential is expected to be linearly rising and the string tension of the quark potential for the representation $r$ is approximately proportional to $C_r$, the eigenvalue of the quadratic Casimir operator for the representation $r$, i.e. $\sigma_{r} \approx\frac{C_{r}}{C_{F}} \sigma_{F}$ where $F$ denotes the fundamental representation \cite{Deldar:1999vi,Bali:2000un,Piccioni:2005un}. When the energy between quarks suffices, a gluon pair is created in the vacuum and Casimir scaling breaks down and is replaced by an $N$-ality dependent law \cite{Kratochvila:2003zj}. Therefore, at asymptotic distances, the quark potential depends on the $N$-ality $k_{r}$ of the representation i.e. $\sigma_{r} = \sigma(k_{r})$. The string tension $\sigma(k_{r})$ corresponds to the lowest dimensional representation of SU($N$) with $N$-ality $k_{r}$. In addition, the lattice results \cite{Bachas:1985xs} show that the static quark potential must be everywhere convex i.e. 
\begin{equation}
  \frac{d V}{d r}>0 \quad \text{and} \quad \frac{d^2 V}{d r^2} \leq 0.
\end{equation}
Therefore, it is crucial to obtain a convex potential without any concavity when interpolating between the short
distances and the asymptotic regimes.

 Any model of the quark confinement should be able to explain these features for the potentials between static quarks. The thick center vortex model has been fairly successful in describing the mechanism of confinement in QCD \cite{Faber:1997rp}. However there are still some shortcomings within the model which is at the focus
of this article. In this model, the vacuum is assumed to be filled with center vortices. In $SU(N)$ gauge group, there are $N-1$ types of center vortices corresponding to 
the nontrivial center elements of $z_n=\exp(i2\pi n/N)\in \mathbb{Z}(N)$ enumerated by the value $n=1,...,N-1$.
The effect of a thick center vortex on a planar Wilson loop is to multiply the loop by a group factor
\begin{equation}
\label{group factor}
W_r(C) \to~ {G}_r(\alpha^{n}_C(x)) W_r(C),
\end{equation}
where the function ${G}_r(\alpha^{n}_C(x))={1}/{d_r}\text{Tr}~\exp [i\vec{\alpha}^{n}_C\vec{{H}}]$, $d_r$ is the dimension of the representation, and $\{H_i\}$ is the set of generators from the Cartan subalgebra. The function $\alpha^{n}_C(x)$ denotes the vortex profile and this angle depends on both the Wilson loop $C$ and the position of the vortex center $x$. If the center vortex is all contained within the Wilson loop $\exp [i\vec{\alpha}^{n}_C\vec{{H}}]=(z_n)^{k_{r}} \mathbb{I}$ where $k_{r}$ is the $N$-ality of representation $r$. Using this constraint, the maximum value of the angle ${\alpha}^{n}_{max}$ could be calculated. If the center vortex is outside the loop $\exp [i\vec{\alpha}^{n}_C\vec{{H}}]=\mathbb{I}$ and therefore it has no
effect on the loop.
The quark potential induced by the center vortices is as follows \cite{Faber:1997rp}:
\begin{equation}
\label{potential}
V_r(R) = -\sum_{x}\ln( 1 - \sum^{N-1}_{n=1} f_{n}
[1 - {\mathrm {Re}}{G}_{r} (\vec{\alpha}^n_{C}(x))]),
\end{equation}
where the parameter $f_n$ determines the probability that any given plaquette is pierced by an nth center vortex. An ansatz for the angle $\vec{\alpha}^{n}_C$ was introduced by Greensite $\it{et~ al.}$ \cite{Greensite:2006sm}. Each
center vortex with square cross section $A_v=L_v \times L_v$ contains small independently fluctuating subregions of area $l^2\ll A_v$ which $l$ is a short correlation length. The only constraint is that the total magnetic fluxes of the subregions 
must correspond to a center element of the gauge group. This square ansatz is as follows:
\begin{equation}
\label{Sansax}
         \vec{\alpha}^n_C(x)\cdot\vec{\alpha}^n_C(x) = \frac{A_v}{ 2\mu} \left[
\frac{A}{ A_v} - \frac{A^2}{ A_v^2} \right]
                         + \left(\alpha^n_{max} \frac{A}{ A_v}\right)^2,
\end{equation}
where $A$ is the cross section of the center vortex overlapping with the minimal area of the Wilson loop and $\mu$ is a free parameter. 

\begin{figure}[h!]
\centering
\includegraphics[width=0.42\columnwidth]{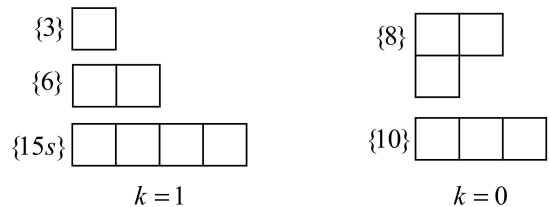}
\caption{The Young diagrams for the lowest representations of SU($3$).
The $N$-ality $k$ of the representations are shown below the diagrams. The label $s$ means the representation is symmetric.}\label{fig:young}
\end{figure}

Now, we apply the model to the SU($3$) gauge group with center $\mathbb{Z}(3)$. The homotopy group
\begin{equation}
\Pi_1[SU(3)/\mathbb{Z}(3)] = \mathbb{Z}(3),
\end{equation}
implies that the SU($3$) gauge theory has center vortices corresponding to the nontrivial
center elements. In SU($3$) case, there are two types of center vortices corresponding to the nontrivial
center elements $z_1=\exp(i2\pi/3)$ and $z_2=\exp(i4\pi/3)$. In some literatures, vortices of type $z_1$ and type $z_2$ have phase
factors which could be considered complex conjugates of one another ($z_1=z_2^*$) and therefore two vortices may be regarded as the
same type of vortex but with magnetic flux pointing in opposite directions. Without this constraint, vortex fluxes of two types of center vortices are different and we analyze the behavior of these center vortices on static potentials. Using Eq. (\ref {potential}), the static potential induced by center vortices in $SU(3)$ gauge group is as follows:
\begin{equation}
V_r(R) =- \sum^{{ {L_v}/2 + R}}_{{x=-{L_v}/2}} \ln[(1-f_1-f_2) + f_1{\mathrm {Re}}{G}_r(\alpha^{1}_C(x))
+ f_2{\mathrm {Re}}{G}_r(\alpha^{2}_C(x))], 
\end{equation}
where $f_1$, $f_2$ are the probabilities that any given plaquette is pierced by $z_1$ and $z_2$ center vortices, respectively. The square ansatz given in Eq. (\ref {Sansax}) for the angles corresponding to the center vortices for all representations are:
\begin{equation}
         (\alpha^1_C(x))^{2} = \frac{A_v}{ 2\mu} \left[
\frac{A}{ A_v} - \frac{A^2}{ A_v^2} \right]
                         + \left(\frac{4\pi}{\sqrt{3}}\frac{A}{ A_v}\right)^2,~~~
 \\
         (\alpha^2_C(x))^{2} = \frac{A_v}{ 2\mu} \left[
\frac{A}{ A_v} - \frac{A^2}{ A_v^2} \right]
                         + \left(\frac{8\pi}{\sqrt{3}}\frac{A}{ A_v}\right)^2.                          
\end{equation}
The free parameters $L_{v}$, $f_{1}$, $f_{2}$, and $L^{2}_{v}/(2\mu)$ 
 are chosen to be $100$, $0.01$, $0.01$, and $4$, respectively. The correlation length is taken $l=1$ and therefore the potentials are linear from the beginning ($R=l$). Now, we study the static potentials of the lowest representations in $SU(3)$ gauge theory. Figure~\ref{fig:young} shows the Young diagrams as well as $N$-ality $k$ of the representations.

\begin{figure}[h!]
\centering
a)\includegraphics[width=0.42\columnwidth]{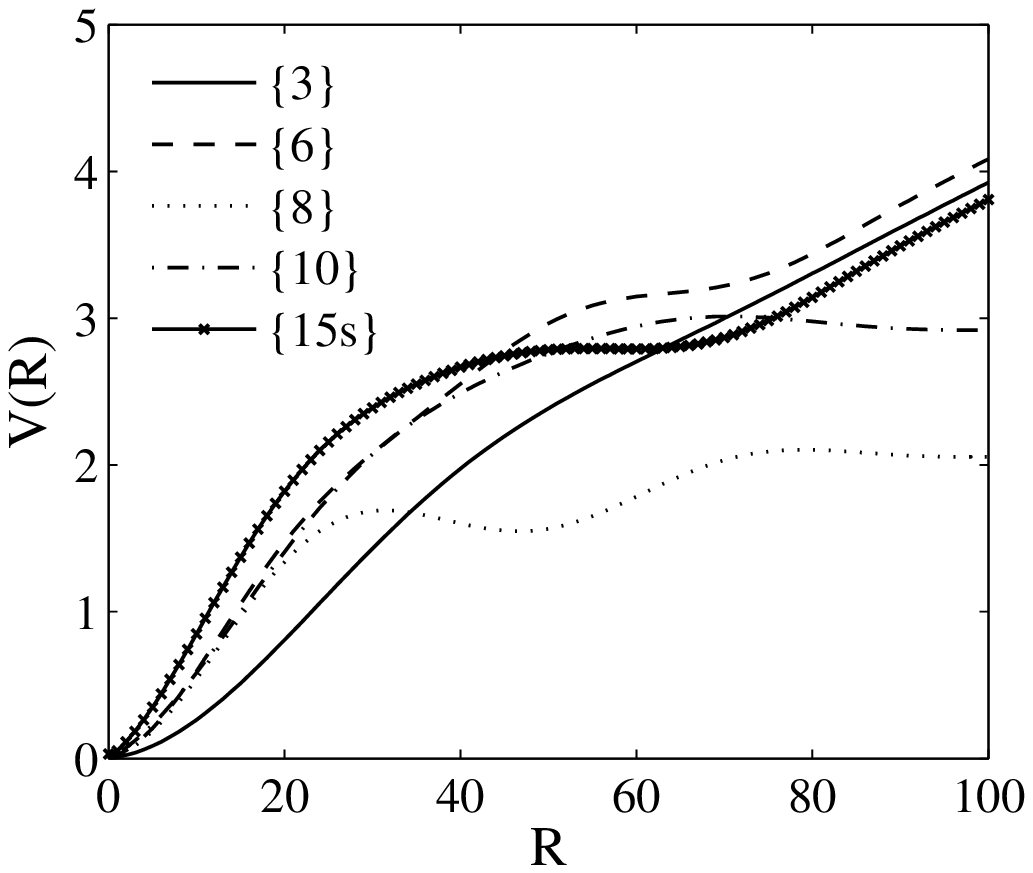}
b)\includegraphics[width=0.42\columnwidth]{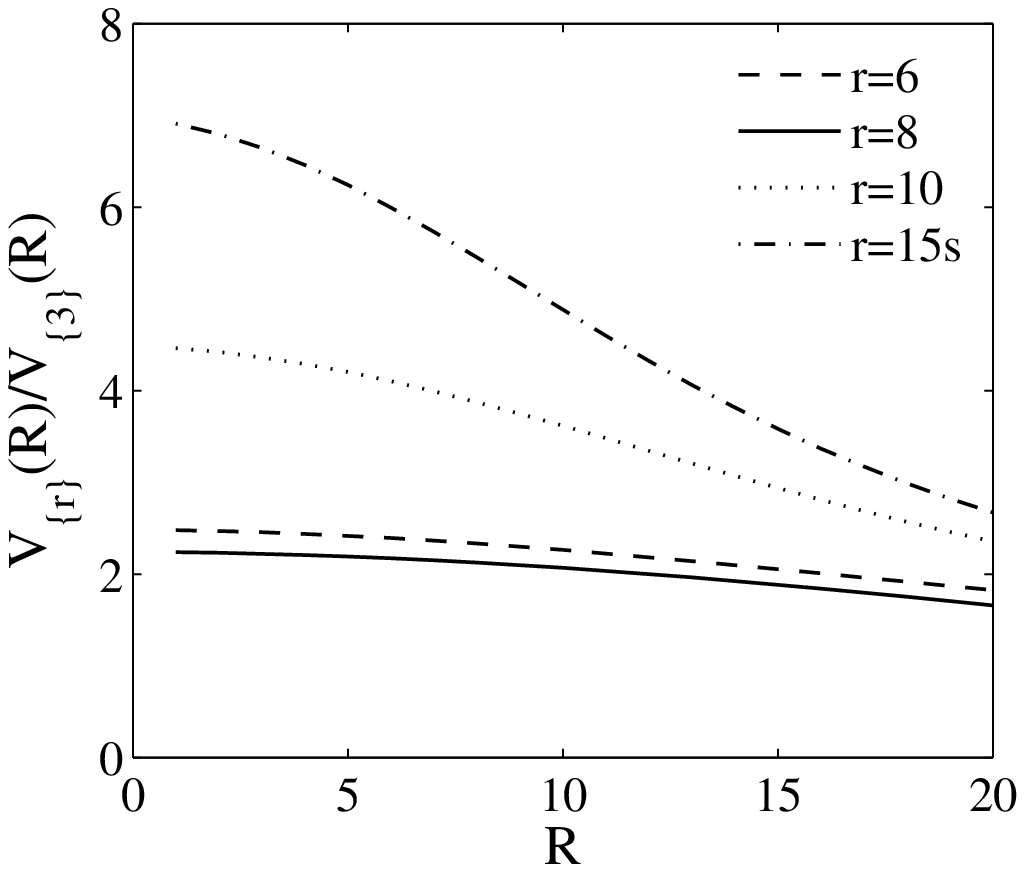}
\caption{a) The static potentials using both types of center vortices for various representations of $SU(3)$. The concavity is appeared for several representations. b) Potential ratios 
$V_{\{r\}}(R)/V_{\{3\}}(R)$ at the intermediate distances. The ratios start from the Casimir ratios and violate slowly from the Casimir ratios in this regime. However, the deviations from the exact Casimir scaling are much greater for higher representations. The free parameters are $L_{v}=100$, $f_{1}=f_{2}=0.01$, and $L^{2}_{v}/(2\mu)=4$.}\label{fig:1}
\end{figure}

Figure \ref{fig:1} a) plots the static potentials $V_{r}(R)$ induced by two types of center vortices for these representations in the range $R\in [0,100]$. At intermediate distances, the potentials are linear in the range $R\in [0,20]$. The potential ratios $V_{\{r\}}(R)/V_{\{3\}}(R)$ for the various representation $r$ are shown in Fig. \ref{fig:1} b). These ratios start out at the Casimir ratios:
\begin{equation}
\frac{C_{\{6\}}}{C_{\{3\}}}=2.5,~~~~~\frac{C_{\{8\}}}{C_{\{3\}}}=2.25,~~~~~ \frac{C_{\{10\}}}{C_{\{3\}}}=4.5,~~~~~ \frac{C_{\{15s\}}}{C_{\{3\}}}=7.
\end{equation}
In the range $R\in [0,20]$, the potential ratios for the various representations drop slowly from Casimir ratios. However, the deviations from the exact Casimir scaling are much greater for higher representations.

At large distances, the static potentials agree with $N$-ality where gluons can bind to
 the initial sources and string tensions of the representations are reduced to the lowest-dimensional representation with the same $N$-ality. In
particular, zero $N$-ality representations are screened. For example, an adjoint charge combining with a gluon can form a color-singlet, $[\{8\} \otimes \{8\}= \{1\} \oplus ...]$. More dynamical gluons might be required for screening of
higher representations with zero $N$-ality. Nonzero $N$-ality representations through combining with gluons are transformed into the
lowest order representations. For example, a tensor product of $[\{6\} \otimes \{8\}=\{\bar{3}\} \oplus ...]$ shows that the slope of the potential for the representation $\{6\}$  must be the same as the one for the representation $\{3\}$. 

As a result, the model leads to Casimir scaling at the intermediate distances and exhibits $N$-ality at the asymptotic regimes, in agreement with lattice calculations. But these two regimes for several representations do not connect smoothly and some kind of unexpected concavity occurs in the model which explicitly disagrees with
lattice results. 

In the next section, for reducing the concavity of some representations, we argue about the behavior of two types of center vortices on the vacuum through analyzing their effects on the Wilson loops.     

\section{Center vortex contributions in the potentials }\label{Sect2}
\begin{figure}[h!]
\centering
\includegraphics[width=0.42\columnwidth]{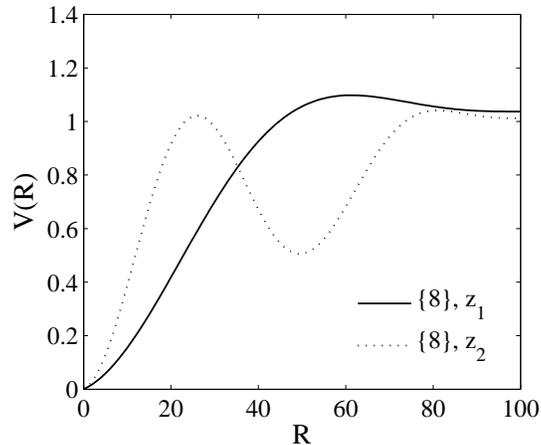}
\caption{ The static potentials induced by two types of center vortices individually for the adjoint representation. The concavity is observed for the potential induced by $z_2$ center vortices  while there is no this artifact in the potential obtained by $z_1$ center vortices. The free parameters are $L_{v}=100$, $f_{1}=f_{2}=0.01$, and $L^{2}_{v}/(2\mu)=4$.}\label{fig:2}
\end{figure}

For analyzing the concavity of the potentials in SU($3$) gauge group, we study the potentials induced by two types of center vortices in more details. As shown in Fig. \ref{fig:1} a), the concavity is appeared for several representations such as the adjoint representation. Figure \ref{fig:2} depicts the potentials induced by two types of center vortices individually for the adjoint representation. The concavity is appeared in the static potential induced by center vortices of type two while there is no this artifact in the potential obtained by center vortices of type one. 

\begin{figure}[h!]
\centering
a)\includegraphics[width=0.43\columnwidth]{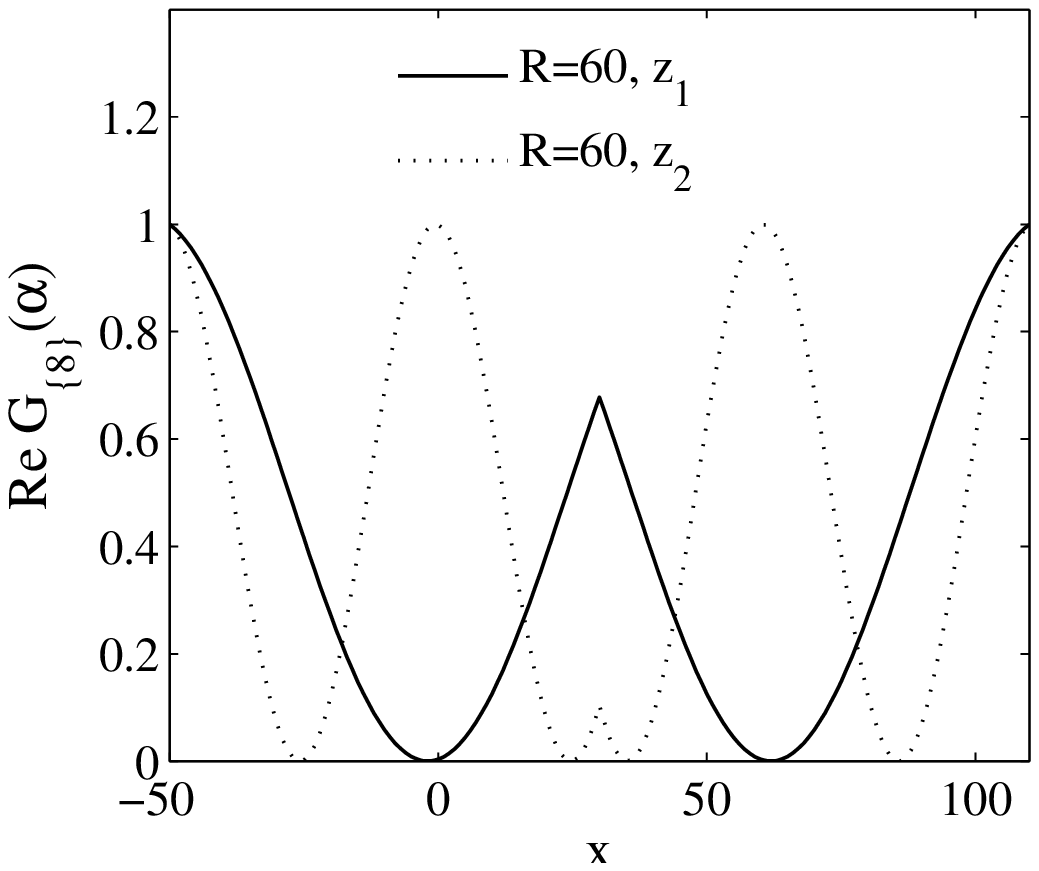}
b)\includegraphics[width=0.42\columnwidth]{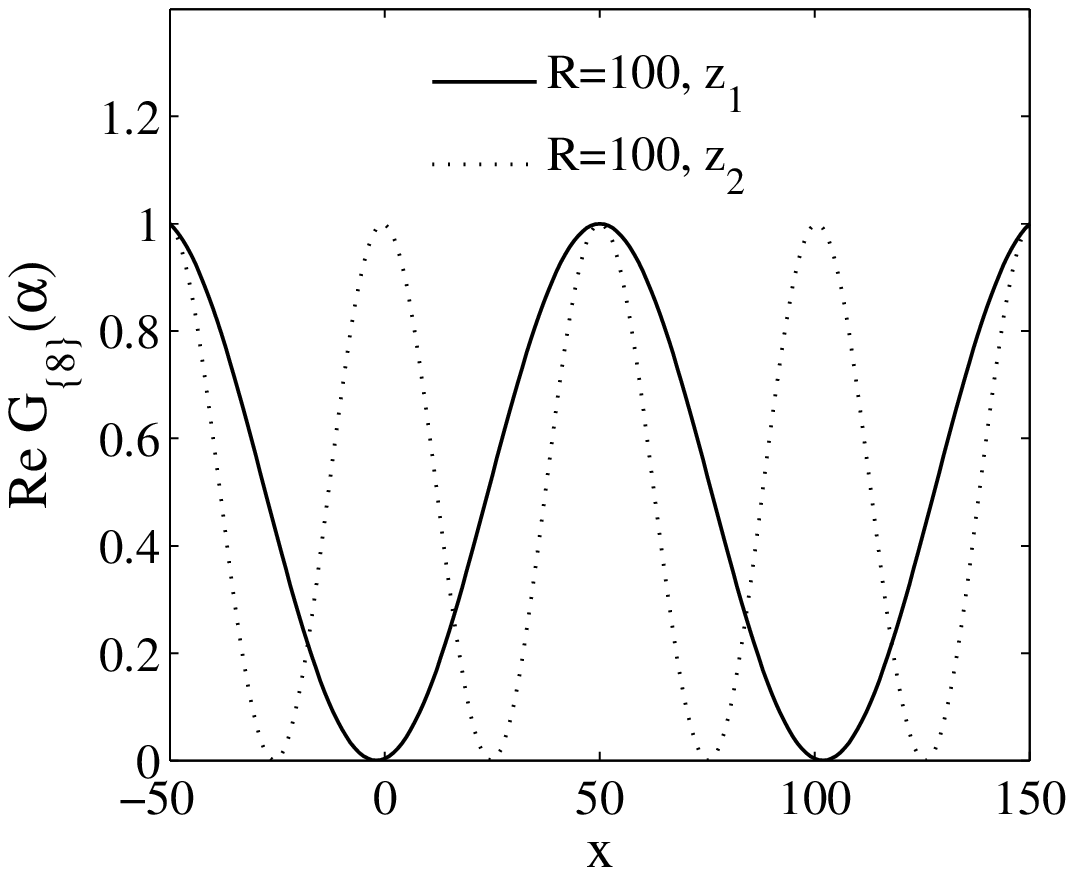}
\caption{ a) The group factors ${\mathrm {Re}}G_{\{8\}}(\alpha)$ of the two types of center vortices versus $x$ corresponding to Fig. \ref{fig:2} at $R=60$, close to the concavity regime. The fluctuations of the $z_2$ vortex group factor with equal large sizes of maxima and minima around any time-like leg lead to concavity behavior in the potentials. b) The same as a) but for the large size Wilson loop with $R=100$. In agreement with the color screening in the large regime for the adjoint representation, the group 
factors for both types of center vortices when the vortex core is located entirely within the Wilson loop is equal to $1$. The free parameters are $L_{v}=100$ and $L^{2}_{v}/(2\mu)=4$.}\label{fig:2-2}
\end{figure}

The behavior of a group factor gives some information about the details of its potential. The functions of the group factors for the lowest representations of SU($3$) can be found in the Appendix. Now, we analyze the group factors corresponding to two types of center vortices for the adjoint representation close to the concavity regime (about $R=60$).  The time-like legs of the Wilson 
loop are located at $x = 0$ and $x = 60$. When the center vortex overlaps the minimal area of the Wilson loop, it affects the Wilson loop. As shown in Fig. \ref{fig:2-2} a), $z_1$ vortex group factor changes smoothly with a minimum value around any time-like leg ($x=0,60$) while for the $z_2$ vortex group factor a wavy character with equal large sizes of maxima and minima is observed the neighborhood of any these regimes. Therefore, the large fluctuations of the group factor around any time-like leg lead to the concavity behavior in the potentials. As shown in Fig.\ref{fig:2-2} b), at large distances ($R=100$) governed with the $N$-ality, the group 
factors for both types of center vortices in the adjoint representation interpolate from $1$, when the vortex core is located entirely within the Wilson loop, to $1$, when the core is entirely outside the loop. As shown in Fig. \ref{fig:2-2}, a fluctuation with a minimum is appeared around each time-like leg for the $z_1$ vortex group factor in the adjoint representation while two of these fluctuations occur around each time-like leg for the $z_2$ vortex group factor. Since $z_2$ vortices are characterized by the center element $z_2=z_1^2$, there is periodicity in the $z_2$ vortex group factor and its potential compared with those of the $z_1$ vortex. 

\begin{figure}[h!]
\centering
a)\includegraphics[width=0.42\columnwidth]{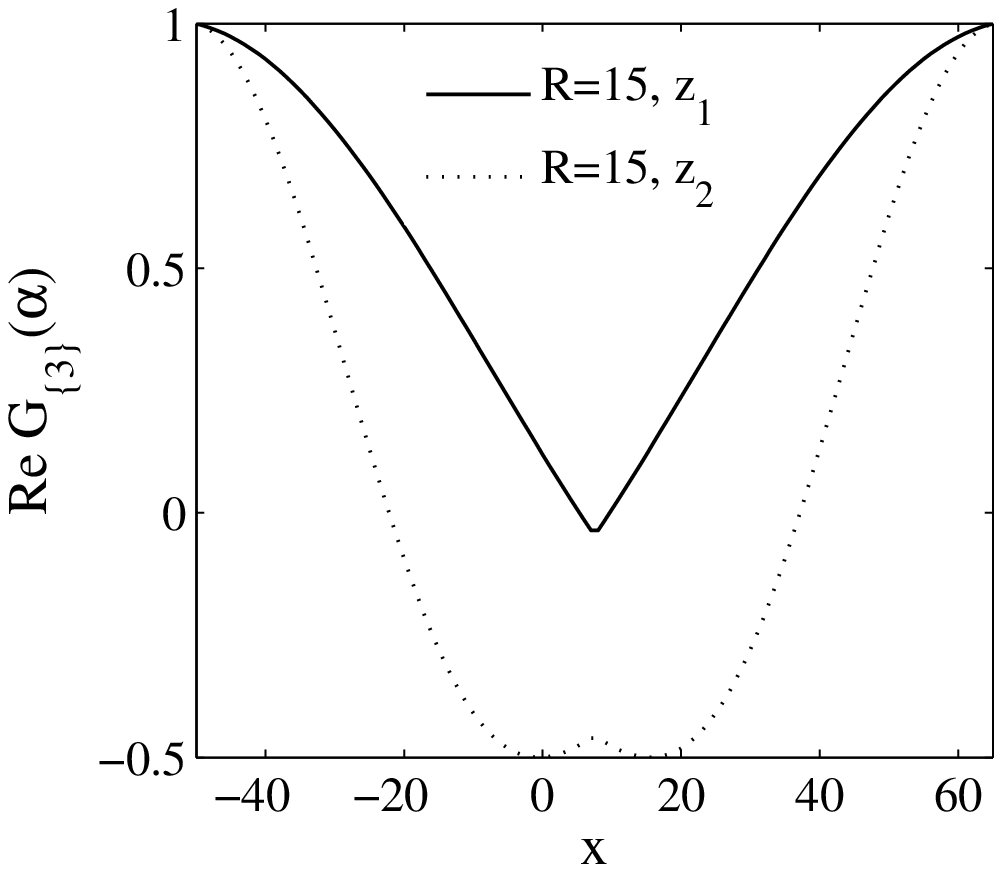}
b)\includegraphics[width=0.42\columnwidth]{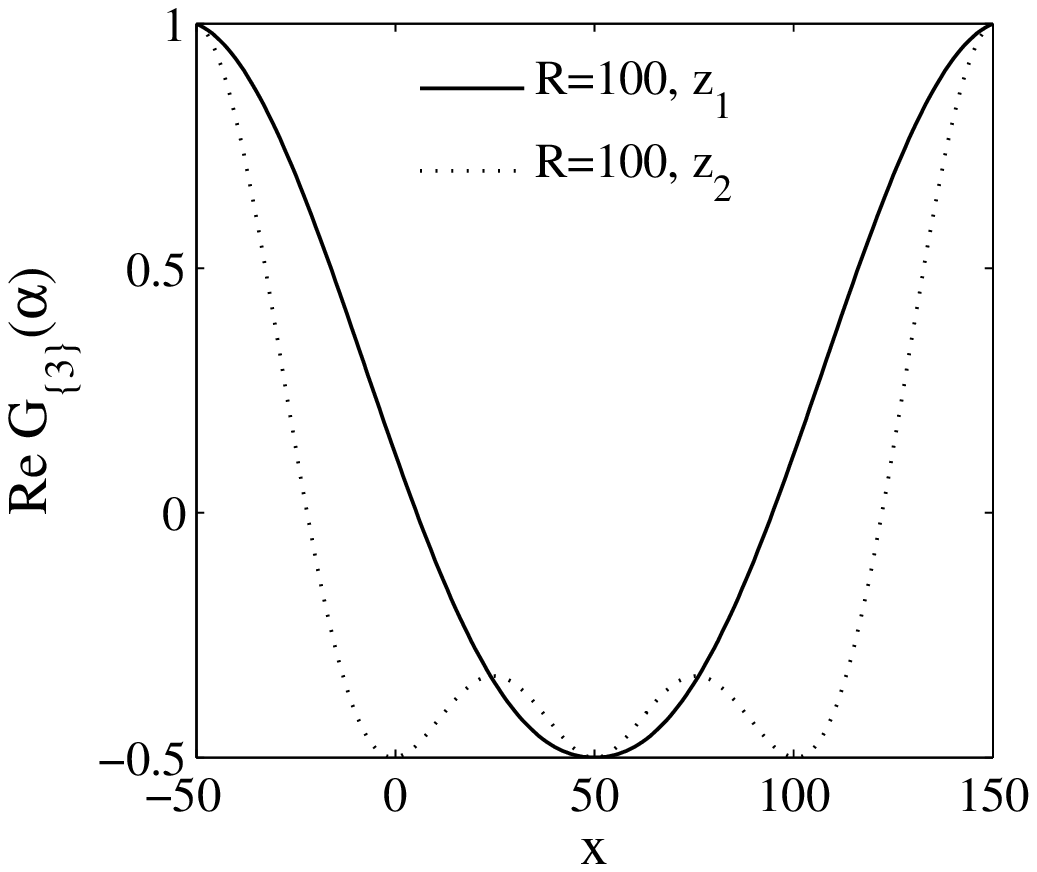}
\caption{a) The group factors ${\mathrm {Re}}G_{\{3\}}(\alpha)$ of the two types of center vortices versus $x$ for the fundamental representation in the intermediate distance with $R=15$. b) the same as a) but for the asymptotic distance with $R=100$. The group 
factors for both types of center vortices when the vortex core is located entirely within the Wilson loop is equal to $-0.5$. Decreasing the size of the loop, the minimum value of the $z_1$ vortex group factor is increased and becomes close to trivial value while the one of the $z_2$ vortex group 
factor is close to center vortex value ($-0.5$). The free parameters are $L_{v}=100$ and $L^{2}_{v}/(2\mu)=4$.}\label{fig:3-1}
\end{figure}

Furthermore, Fig. \ref{fig:3-1} a) depicts the group factors corresponding to two types of center vortices for the medium size Wilson loop with $R=15$ for the fundamental representation and the ones for the large size loop with $R=100$ are plotted in Fig. \ref{fig:3-1} b). As shown, the $z_1$ vortex group factor in the fundamental representation changes smoothly around each time-like leg and therefore one could expect that the group factor of $z_2=z_1^2$ changes smoothly the neighborhood of each time-like leg. 

\begin{figure}[h!]
\centering
a)\includegraphics[width=0.42\columnwidth]{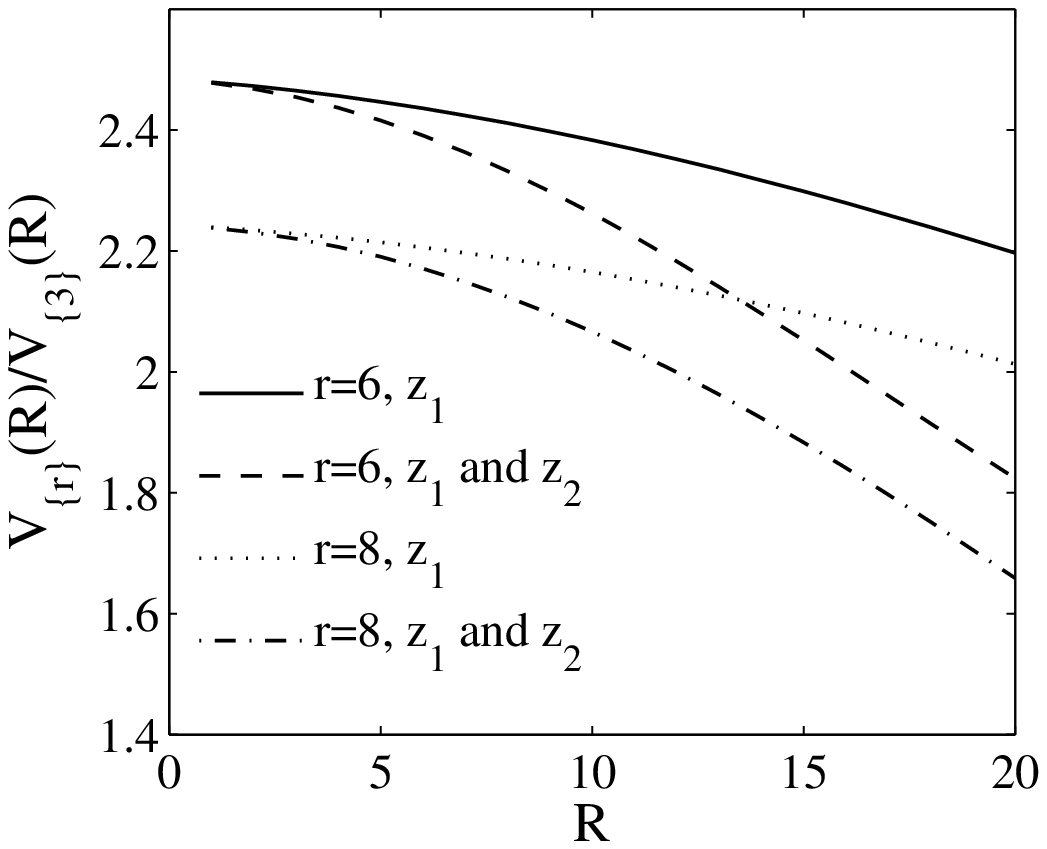}
b)\includegraphics[width=0.42\columnwidth]{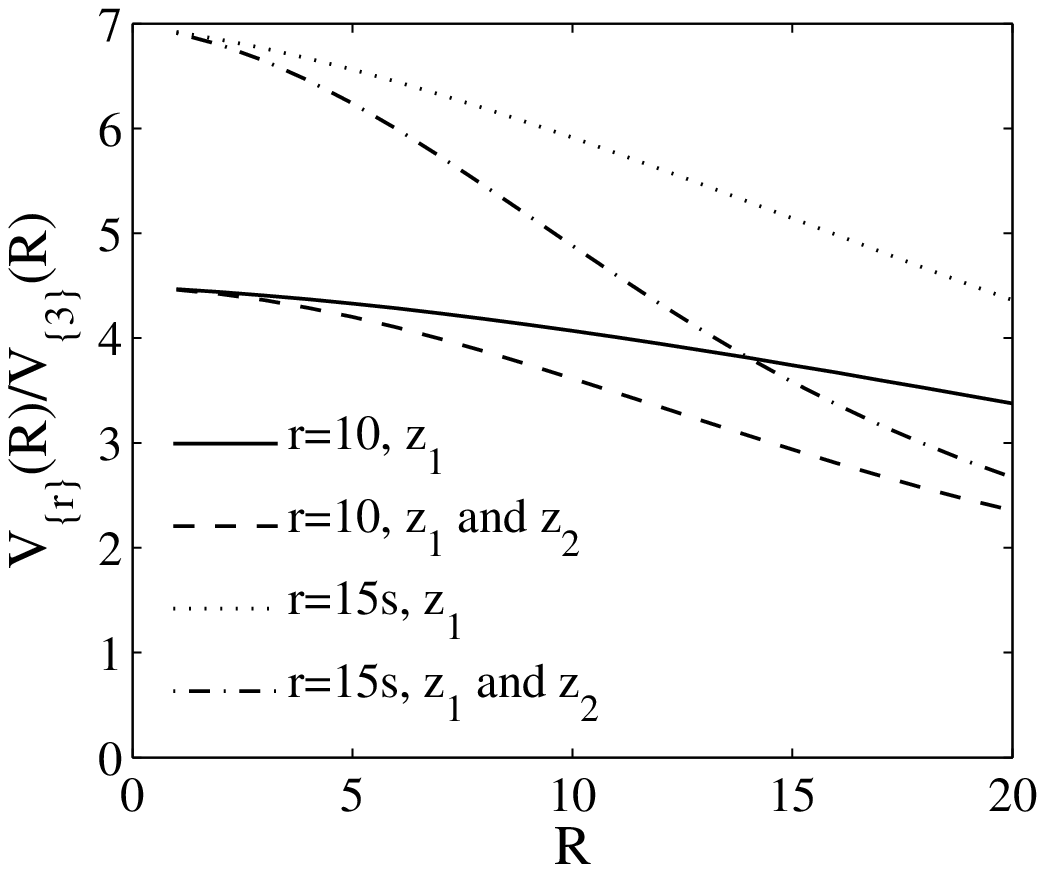}
\caption{The potential ratios of ${V_{\{r\}}(R)}/{V_{\{3\}}(R)}$ induced by center vortices for the various representations. Upper curve of any representation shows the contribution of the $z_1$ vortices which violates more slowly from the Casimir 
ratio compared with the contribution of the $z_1$ vortices plus the $z_2$ vortices. For all representations, the ratios induced by the $z_1$ vortices agree better with Casimir scaling compared
with  the ratios induced by  $z_2$ vortices. The free parameters are $L_{v}=100$, $f_{1}=f_{2}=0.01$, and $L^{2}_{v}/(2\mu)=4$.}\label{fig:3}
\end{figure}

At large distances, the group 
factors for both types of center vortices in the fundamental representation interpolate from $-0.5$, when the vortex core is located entirely within the Wilson loop, to $1$, when the core is entirely outside the loop. As shown in Fig. \ref{fig:3-1} a), decreasing the size of the Wilson loop ($R=15$), the minimum value of the $z_1$ vortex group factor is increased while the one of the $z_2$ vortex group 
factor is close to center vortex value ($-0.5$). The value of the $z_2$ vortex group factor for the medium size Wilson loops is about center vortex value which is related to $N$-ality regimes. Therefore, we expect that the $z_2$ vortices break down somewhat the Casimir scaling at intermediate distances.  Figure \ref{fig:3} plots the potential ratios induced by center vortices for the range $R\in [0,20]$. As shown, the contributions of two types of the center vortices are compared.  For various representations, the potential ratios obtained from the $z_1$ vortices which start from the Casimir ratios drop slower than those induced by both types of vortices.

\begin{figure}[h!]
\centering
\includegraphics[width=0.42\columnwidth]{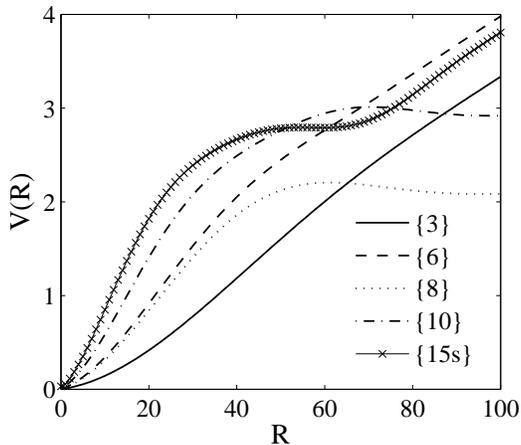}
\caption{The static potentials using $z_1$ center vortices for the various representations of $SU(3)$. Although the concavity is removed somewhat this artifact stays for higher presentations especially $\{15s\}$. The free parameters are $L_{v}=100$, $f_{1}=0.01$, and $L^{2}_{v}/(2\mu)=4$.}\label{fig:5}
\end{figure}

For detailed analysis of two types of vortices, we note to the interactions between vortices.  The QCD vacuum could be described in terms of a
Landau-Ginzburg model of a dual superconductor where it follows the electric flux tube formation and confinement of the electric charge. A dual superconductor is like type II superconductors but the roles of the electric and magnetic
fields, and electric and magnetic charges, have been interchanged \cite{Hooft,Mandelstam}. Properties of superconductors are often described in terms of the superconducting coherence length $\displaystyle \xi$ and the London magnetic field penetration depth $\displaystyle \lambda$. The Ginzburg-Landau
parameter $\kappa=\lambda /\xi$ of the type-II superconductor is larger than $1/\sqrt{2}$. In the type II superconductors, there are vortices as the magnetic flux lines as well as the magnetic fluxes pointing in opposite directions of vortices (antivortices). The interaction between vortices is repulsive while the vortex-antivortex interaction is attractive \cite{Kramer:1971zza,Chaves}. Furthermore, one may find the same interactions between vortices in the model which is discussed in the next section. Now, using these results, the vacuum is argued in SU($3$) case, filled with $z_2$ vortices as well as $z_1$ vortices. Such $z_2$ vortices are characterized by the center element $z_2=z_1^2$. The $z_2$ vortex is constructed of two $z_1$ vortices with the same flux orientations and therefore these $z_1$ vortices according to the interactions in the type-II superconductor repel each other. One may conclude that $z_2$ vortices do not make a stable configuration and one should consider each of $z_1$ vortices within the $z_2$  vortices as a single vortex
in the model. In addition, only vortices with the smallest magnitude of center flux have substantial probability \cite{Faber:1997rp}.
 In fact, this probability for the $z_2$ vortex should be less than the one for the $z_1$ vortex. In previous section, we considered the general case that all possible $f_n$ are included and therefore the concavity is appeared for several representations. Now, we assume only ${z_1}$ vortices in the vacuum. Figure \ref{fig:5} shows the potentials for the various representations in the range $R\in [0,100]$. 

Although, using only ${z_1}$ vortices, two Casimir scaling and $N$-ality distances are smoothly connected for some representations, the concavity occurs for higher representations especially $\{15s\}$. Indeed, this concavity is observed independent of the ansatz for the angle \cite{Deldar:2010hw}.  

The next step, the confinement mechanism in the background of center vortices would be reformulated for removing this concavity and we discuss two types of SU($3$) center vortices in more details. 

\section{Vacuum domains and removing the concavity of the potentials }\label{Sect3}

The QCD vacuum is a dual analogy to the type II superconductivity. As argued, it seems that two vortices repel each other while the vortex-antivortex interaction is attractive. On the one hand, two $z_1$ vortices within the $z_2$ vortex repel each other and one could observe them as the single vortices. On the other hand, in addition to two types of center vortices, there are their antivortices corresponding to complex conjugates of center elements on the vacuum. Therefore, $z_2$ and $z_1^*$ vortex configurations attract each other and they would merge forming $z_2z_1^*=z_1^2z_1^*=z_1 z_0$ where $z_1z_1^*$ is equal to the identity element $z_0=1$. In Refs. \cite{Greensite:2006sm,Liptak:2008gx,Nejad:2014tja}, vacuum domains corresponding to the identity element are also allowed in the model. The Yang–Mills vacuum has a domain structure where there are domains of the center-vortex type and of the vacuum type. Therefore, we observe that attractions between $z_2$ and $z_1^*$ vortices are forming $z_1$ vortices as well as vacuum domains. One could apply the same argument for $z_2^*$ and $z_1$ vortex configurations. The domain structures can be readily generalized to SU($4$) and beyond. For example, in SU($4$), there are non-trivial center elements $z_1=\exp(i\pi/2)$, $z_2=z_1^2$, and $z_3=z_1^3$. The attractions between vortices and anti-vortices in SU($4$) may form center vortices as well as vacuum domains. Using Eq. (\ref {potential}), the static potential induced by center vortices as well as the vacuum domains is:
\begin{equation}
\label{potential2}
V_r(R) = -\sum^{{ {L_v}/2 + R}}_{{x=-{L_v}/2}}\ln( 1 - \sum^{N-1}_{n=0} f_{n}
[1 - {\mathrm {Re}}{G}_{r} (\vec{\alpha}^n_{C}(x))]),
\end{equation}
where the contribution of the vacuum domains ($n=0$) is added. If a vacuum domain is all contained within the Wilson loop $\exp [i\vec{\alpha}^{0}_C\vec{{H}}]=z_0\mathbb{I}$. The total magnetic flux through a vacuum domain is zero value and therefore the square ansatz given in Eq. (\ref {Sansax}) for the angle of vacuum domain for all representations is
\begin{equation}
         (\alpha^0_C(x))^{2} = \frac{A_v}{ 2\mu} \left[
\frac{A}{ A_v} - \frac{A^2}{ A_v^2} \right].                          
\end{equation}.

\begin{figure}[h!]
\centering
\includegraphics[width=0.42\columnwidth]{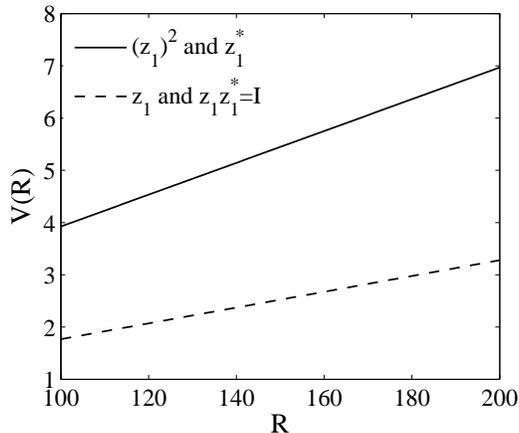}
\caption{The static potential induced by $z_1$ and $z_0$ vortex configurations is compared with the one induced by $z_1^*$ and $z_2$ vortex configurations in the fundamental representation at large distances. It seems that for minimizing the energy $z_2$ and $z_1^*$ attract each other and deform to $z_1$ and $z_0$ vortex configurations. For each configuration, the free parameters are $f_n=0.01$, $L_{v}=100$, and $L^{2}_{v}/(2\mu)=4$.}\label{fig:4}
\end{figure}

Now, to understand the interactions between two types of SU($3$) center vortices, we study the static potentials in the fundamental representation. Figure \ref{fig:4} shows the static potentials induced by vortex configurations in the fundamental representation at large distances where the ansatz of the vortex profile has no role in the potentials. Each configuration is appeared in the plane of the Wilson loop with the probability $f_n=0.01$. We assume that there are $z_1$ vortex as well as $z_1^*$ antivortex on the vacuum. Adding $z_1$ vortex to the vortex configurations may lead either to $z_2=z_1^2$ and $z_1^*$ vortex configurations or to $z_1$ and $z_0=z_1^*z_1$ vortex configurations.  
In other words, it is interesting to observe that this $z_1$ vortex is attracted by which one of the initial vortices, $z_1$ vortex or $z_1^*$ antivortex. One expects that the ensemble of the vortex configurations leads to a minimum energy. The potential energy induced by $z_1^*$ and $z_2$ vortex configurations is more than the one induced by $z_1$ and $z_0$ vortex configurations. It seems that for minimizing the energy $z_2$ and $z_1^*$ attract each other and deform to $z_1$ and $z_0$ vortex configurations. The extra negative energy of the potential induced by $z_1$ and $z_0$ vortex configurations compared with the one induced by $z_1^*$ and $z_2$ vortex configurations may be interpreted as an attraction energy between $z_1$ and $z_1^*$ vortices and repulsion between two $z_1$ vortices. Therefore, two magnetic vortex fluxes with the same orientation may repel each other while those with the opposite orientation attract each other. It seems that the model also confirms the interactions between vortices in the type II superconductivity. In addition, we studied the interaction between vortices in the model based on energetics in Refs. \cite{Nejad:2017njp,Nejad:2014tja} approving the same results for the interactions. It seems that the attractions between two types of center vortices produce vortices of type $n=1$ and vacuum domains. It is possible that only vortices with the smallest magnitude of center flux have substantial probability to find the midpoints
of them at any given location \cite{Faber:1997rp}. It seems that $z_2$ vortices, which its magnitude of center flux is twice the one of $z_1$ vortices, interacting with $z_1$ vortices are decomposed to the configurations with the lowest magnitude of center fluxes, i.e., $z_1$ vortices and vacuum domains. 

\begin{figure}[h!]
\centering
\includegraphics[width=0.42\columnwidth]{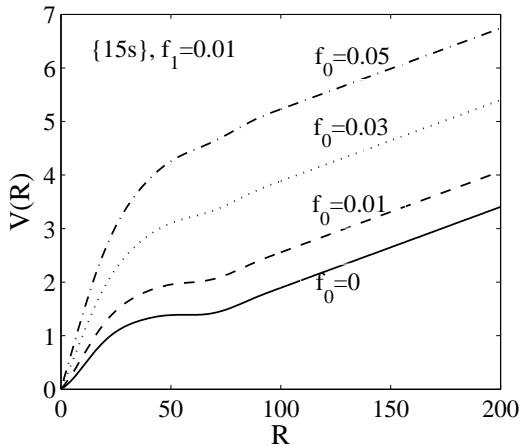}
\caption{The static potential for the representation $\{15s\}$. On the vacuum, there are $z_1$ vortices with the fixed probability $f_1=0.01$ while the probability $f_0$ of vacuum domains is gradually increased. The concavity could almost be removed by appearing the vacuum domains, a way to effectively parametrize vortex interactions. The free parameters are $L_{v}=100$ and $L^{2}_{v}/(2\mu)=4$.}\label{fig:6}
\end{figure}

Therefore, in SU($3$) case using Eq. (\ref {potential2}), the static potential induced by $z_1$ vortices and vacuum domains is
\begin{equation}
V_r(R) =- \sum^{{ {L_v}/2 + R}}_{{x=-{L_v}/2}} \ln[(1-f_0-2f_1) +f_0{\mathrm {Re}}{G}_r(\alpha^{0}_C(x))+ 2f_1{\mathrm {Re}}{G}_r(\alpha^{1}_C(x)) ], 
\end{equation}
where the vortices of type $n=2$ is substituted with those of type $n=1$ and vacuum domains. $f_0$, $f_1$ are the probabilities that any given plaquette is pierced by vacuum domains and $z_1$ vortices, respectively. In Fig. \ref{fig:6}, the static potential for the representation $\{15s\}$ which has shown the worst concavity is plotted. On the vacuum, there are $z_1$ vortices with the fixed probability $f_1=0.01$ but the probability $f_0$ of vacuum domains is gradually increased from zero to $0.05$. As shown, the concavity could almost be removed by appearing the vacuum domains in the vacuum. The concavity for the higher representation $\{27\}$ is also eliminated. Therefore, the satisfactory result can be achieved by ad-hoc choosing the probability weights of the different vortex configurations. In particular including vacuum domains, a way to effectively parametrize
vortex interactions, is crucial in obtaining an (almost)
everywhere convex potential.

\begin{figure}[h!]
\centering
a)\includegraphics[width=0.42\columnwidth]{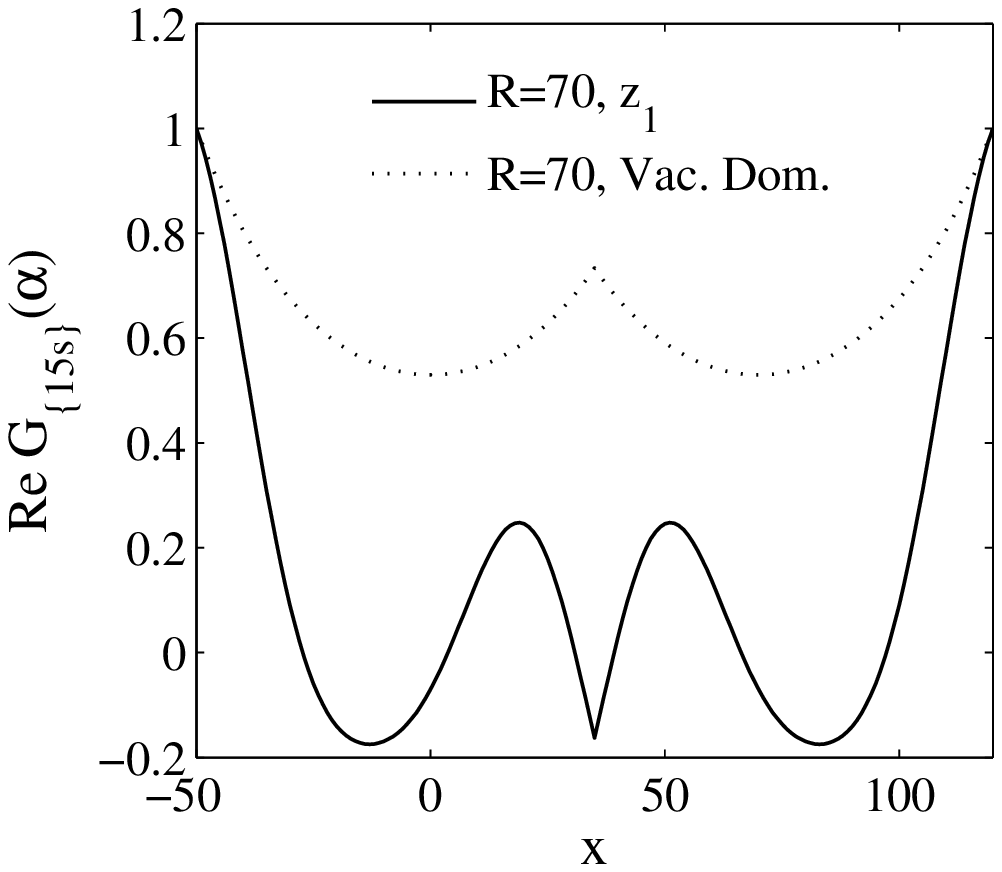}
b)\includegraphics[width=0.42\columnwidth]{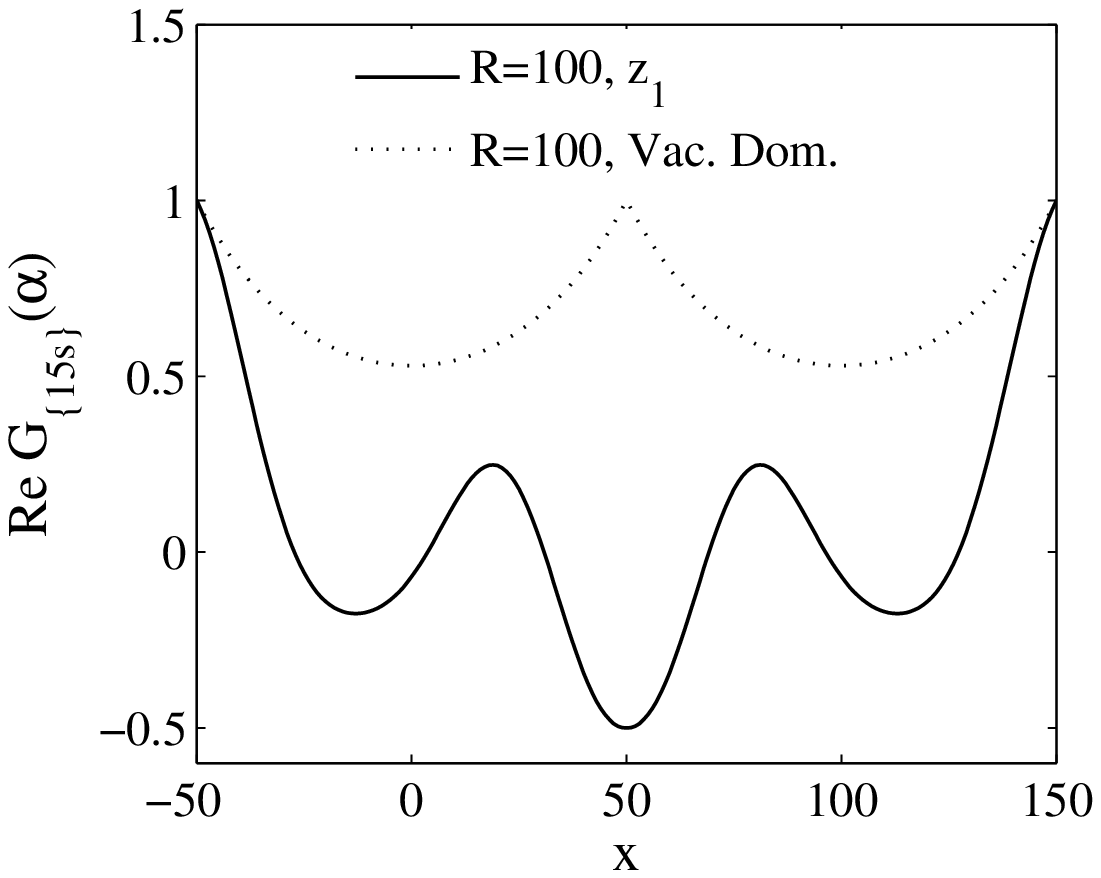}
\caption{a) The group factors ${\mathrm {Re}}G_{\{15s\}}(\alpha)$ of the $z_1$ vortices and vacuum domains versus $x$ at $R=70$ corresponding to Fig. \ref{fig:6}, close to the concavity regime. The fluctuations of the $z_1$ vortex group factor with equal large sizes of maxima and minima around any time-like leg lead to the concavity behavior in the potentials. The vacuum domain group factor changes smoothly close to the trivial value $1$ around any time-like leg removing the concavity in the potential. b) The same as a) but for the large size Wilson loop with $R=100$. Since the $N$-ality of the representation {\{15s\}} is the same as the one of the fundamental representation, the $z_1$ vortex group factor like the one of the fundamental representation interpolates from $-0.5$, when the vortex core is located entirely within the Wilson loop, to $1$, when the core is entirely outside the loop. Also, in the same interval, when the core of vacuum domain is located entirely within the loop, the group 
factor reaches to the trivial value $1$. The free parameters are $L_{v}=100$ and $L^{2}_{v}/(2\mu)=4$.}\label{fig:6-2}
\end{figure}
To check the details of the static potential in the representation $\{15s\}$, we analyze its group factors. Figure \ref{fig:6-2} a) depicts the group factors corresponding to $z_1$ vortices and vacuum domains for the medium size Wilson loop close to concavity regime (about $R=70$) for the representation $\{15s\}$ and those for the large size loop are plotted in Fig. \ref{fig:6-2} b). As shown in Fig. \ref{fig:6-2} a), for the $z_1$ vortex group factor, a wavy character with a large amplitude appears around to any time-like leg ($x=0,70$). The same behavior occurs for the $z_2$ vortex group factor in the adjoint representation and therefore the concavity is appeared in its potential. Increasing large fluctuations of the group factor leads to the concavity behavior in the potentials. But, the vacuum domain group factor changes smoothly (small fluctuations) close to trivial value $1$ around any time-like leg. Therefore the concavity of the potential could be removed by including the vacuum domain contribution to the potential. As shown in Fig. \ref{fig:6-2} b), the group 
factor for $z_1$ vortices in the representation $\{15s\}$ at large distances ($R=100$), like the one of the fundamental representation, interpolates from $-0.5$, when the vortex core is located entirely within the Wilson loop, to $1$, when the core is entirely outside the loop. Also, in the same interval, when the core of the vacuum domain is located entirely within the loop, the group 
factor reaches to the trivial value $1$. Besides, as show in Ref. \cite{Nejad:2014tja}, the vacuum domains could enhance the Casimir scaling at the intermediate distances.

 As a result, small fluctuations of the group factor close to the trivial value, which occur because of the interactions between center vortices, could remove concavity in the static potentials and also improve the Casimir scaling at the intermediate regime. But the large fluctuations of the group factor could create the concavity in the static potentials and break down the Casimir scaling at the intermediate regime. 

\begin{figure}[h!]
\centering
a)\includegraphics[width=0.42\columnwidth]{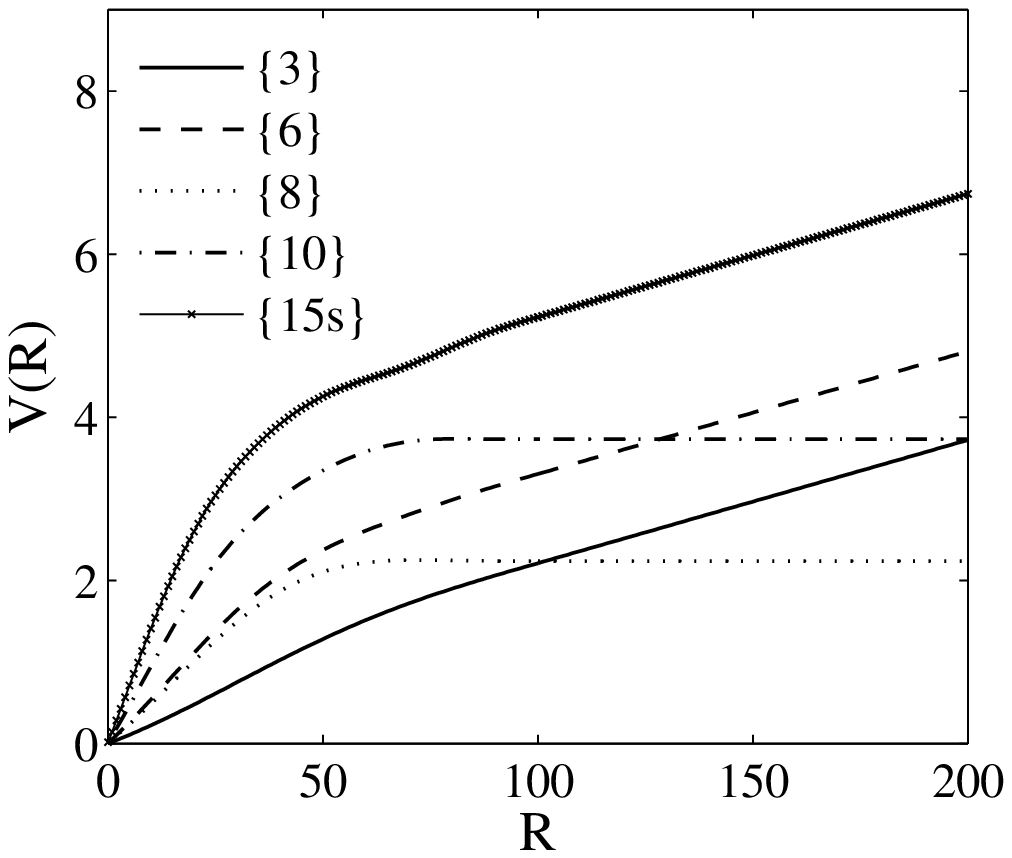}
b)\includegraphics[width=0.42\columnwidth]{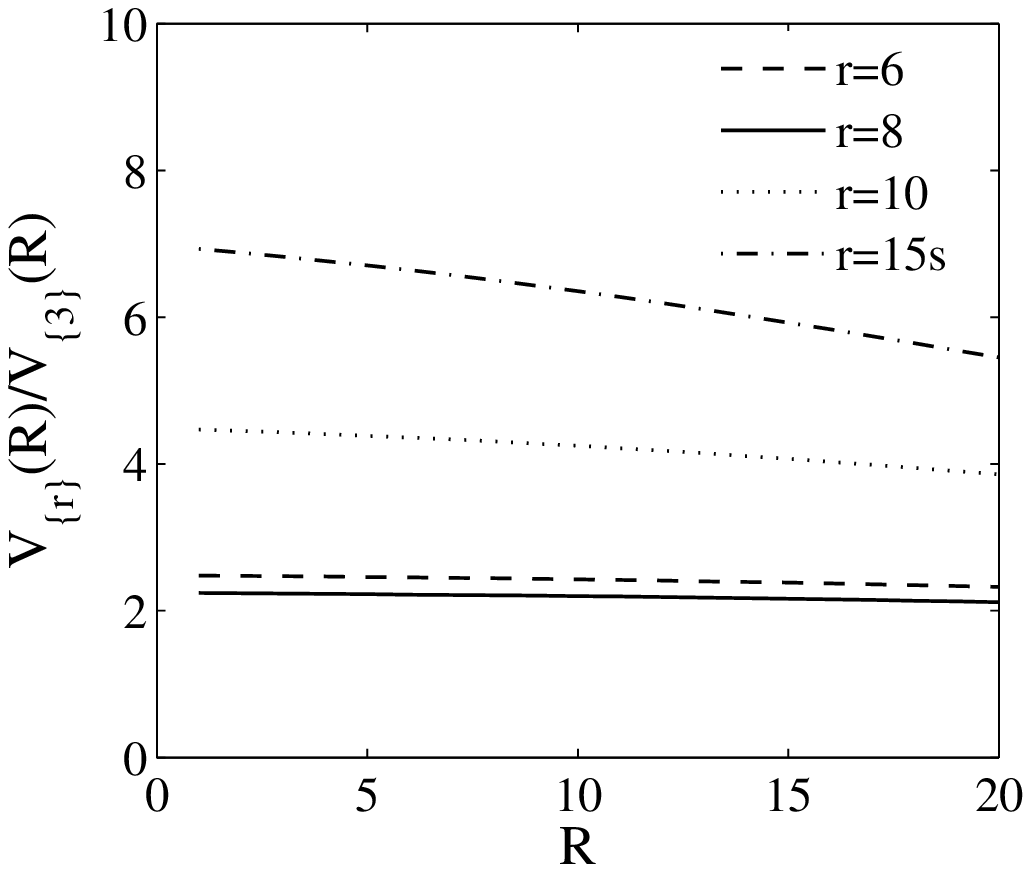}
\caption{a) The static potentials using both $z_1$ center vortices and vacuum domains for the various representations of $SU(3)$. The Casimir scaling and $N$-ality regimes connect naturally to each other without almost any concavity. b) Potential ratios 
$V_{\{r\}}(R)/V_{\{3\}}(R)$ at the intermediate distances. These potentials agree with Casimir scaling better than those obtained from both types of vortices, especially for the higher representations. The free parameters are $L_{v}=100$, $f_{0}=0.05$, $2f_{1}=0.01$, and $L^{2}_{v}/(2\mu)=4$.}\label{fig:7}
\end{figure}

In Fig. \ref{fig:7} a), the potentials $V_{r}(R)$ induced by $z_1$ vortices and vacuum domains for the various representations for the range $R\in [0,200]$ are plotted and the potential ratios are shown in Fig. \ref{fig:7} b). Therefore, the satisfactory potentials for
the different representations can be achieved by ad-hoc choosing the probability weights of the
different vortex configurations. In particular including the vacuum domains, a way to effectively parametrize
vortex interactions, is crucial in obtaining an (almost) everywhere convex potential when interpolating between the short distances and the asymptotic regimes. In addition, the potential ratios starting out at the Casimir ratios at intermediate distances drop very slowly from the exact Casimir scaling for all representations, especially for the higher representations. Therefore, the convex potentials in agreement with Casimir scaling at intermediate regimes with a fixed vortex profile could be obtained, if one includes the contribution of vortex interactions in the static potentials. 

Furthermore, when the properties of vortices in $d=2$ dimensions and $\mathbb{Z}(N)$
models were being worked out, it was found that a real-space
renormalization group approach to the $\mathbb{Z}(N)$ models reproduced the
correct change in critical behavior at $N=4$ if vacancies were included.
In fact, it was found that the vacancy fugacity mimicked the vortex
fugacity, and was a relevant variable in the disordered phase. In the framework of the real-space renormalization group approach \cite{Nienhuis,Buessen,Canet}, analyzing the convexity could be interesting and we will focus on this idea in the future works.

\section{Conclusion}\label{Sect4}
The static potentials in various representations depend on basic properties. At the intermediate regime, the potentials are governed by Casimir scaling while this feature breaks down in the asymptotic regime and is replaced by the $N$-ality dependent law. These two regimes should be connected smoothly to each other without any concavity. In this paper, we analyze the static potentials in SU($3$) Yang-Mills theory
within the framework of the domain model of center vortices where there are two types of vortices. The two types of vortices may be regarded as the same type of vortex with magnetic flux pointing in opposite directions. Without this constraint, we study the behavior of these center vortices on static potentials. The interactions of both types of center vortices with large size Wilson loops are the same but their interactions with the medium size Wilson loops are different. The potentials induced by both vortex types show concave behavior for several representations. In addition, the potential ratios induced by $z_1$ vortices starting out at the Casimir ratios at intermediate distances drop slower than those of $z_2$ vortices. Analyzing the interactions between two types of center vortices, the confinement mechanism of center vortices is reformulated for removing the concavity of the potentials and also improving the Casimir scaling at the intermediate regimes. The QCD vacuum is a dual analogy to the type
II superconductivity where it seems that two vortices repel each other while the vortex-antivortex interaction is attractive. We show that the model may also confirm the same interactions between vortices based on energetics.  On the one hand, $z_2$ vortices are characterized by the center element $z_2=z_1^2$ and two $z_1$ vortices within the $z_2$ vortex may repel each other and one could observe them as the single vortices. However using only ${z_1}$ vortices, this concavity would still remain for some higher representations. On the other hand, in addition to two types of center vortices, there are their antivortices on the vacuum. We show, like superconductivity, that $z_2$ and $z_1^*$ vortex configurations may attract each other and therefore they would merge forming $z_2z_1^*=z_1z_0$ where $z_0$ is equal to the identity element. We observe that attractions between $z_2$ and $z_1^*$ vortices are forming $z_1$ vortices as well as vacuum domains. Therefore, $z_2$ vortices, which its magnitude of center flux is twice the one of $z_1$ vortices, within the interactions with the $z_1$ vortices may be decomposed to the configurations with the lowest magnitude of center fluxes. As a result, the vacuum in stead of $z_1$ and $z_2$ vortices is filled with $z_1$ vortices and vacuum domains. We show that by ad-hoc choosing the probability weights of the
different vortex configurations, satisfactory result for the static potentials can be achieved. In particular including the vacuum domains, a way to effectively parametrize
vortex interactions, is crucial in obtaining the convex potentials in agreement with Casimir scaling at intermediate regimes.

\appendix
\section{ Group factors of the representations}

 The Cartan generators for the representation $r$ within the group factors of the static potential given in Eq. (\ref {potential2}) can be calculated using the tensor method. One can obtain the real part of the group factors for all center domains in several representations as:

\begin{equation}
    {\mathrm {Re}}{G}_{\{3\}}(\alpha^{n})=\frac{1}{3}[2cos(\frac{\alpha^{n}}{2\sqrt{3}})+cos(\frac{\alpha^{n}}{\sqrt{3}})],   
\label{group-3}                      
\end{equation}
\begin{equation}
    {\mathrm {Re}}{G}_{\{6\}}(\alpha^{n})=\frac{1}{6}[2cos(\frac{\alpha^{n}}{2\sqrt{3}})+3cos(\frac{\alpha^{n}}{\sqrt{3}})+cos(\frac{2\alpha^{n}}{\sqrt{3}})],  
\label{group-6}                       
\end{equation}
\begin{equation}
    {\mathrm {Re}}{G}_{\{8\}}(\alpha^{n})=\frac{1}{8}[4+4cos(\frac{3\alpha^{n}}{2\sqrt{3}})],  
\label{group-8}                       
\end{equation}
\begin{equation}
    {\mathrm {Re}}{G}_{\{10\}}(\alpha^{n})=\frac{1}{10}[3+6cos(\frac{3\alpha^{n}}{2\sqrt{3}})+cos(\frac{6\alpha^{n}}{2\sqrt{3}})],   
\label{group-10}                      
\end{equation}
\begin{equation}
    {\mathrm {Re}}{G}_{\{15s\}}(\alpha^{n})=\frac{1}{15}[4cos(\frac{\alpha^{n}}{2\sqrt{3}})+3cos(\frac{\alpha^{n}}{\sqrt{3}})+5cos(\frac{2\alpha^{n}}{\sqrt{3}})+2cos(\frac{5\alpha^{n}}{2\sqrt{3}})+cos(\frac{4\alpha^{n}}{\sqrt{3}})].   
\label{group-15}                      
\end{equation}

%

\end{document}